\documentclass[12pt,preprint]{aastex62}

\shorttitle{JAGB Distance Scale}
\shortauthors{Madore \& Freedman}

\begin{document}


\title{\bf Astrophysical Distance Scale \\ The JAGB Method:  \\ I. ~Calibration and a First Application}


\author{\bf Barry F. Madore} 
\affil{The Observatories \\ Carnegie
Institution for Science \\ 813 Santa Barbara St., Pasadena, CA ~~91101}
\affil{Dept. of Astronomy \& Astrophysics \\ University of Chicago \\ 5640 S. Ellis Ave.,
\\ Chicago, IL, ~~60637} 
\email{barry.f.madore@gmail.com} 

\author{\bf Wendy L. Freedman}
\affil{Dept. of Astronomy \& Astrophysics \\ University of Chicago \\ 5640 S. Ellis Ave.,
\\ Chicago, IL, ~~60637} 
\email{wfreedman@uchicago.edu}


\begin{abstract} 

$J$-Branch Asymptotic Giant Branch (JAGB) stars are a photometrically well-defined 
population of extremely red, intermediate-age AGB stars that are found to have 
tightly-constrained luminosities in the near-infrared.  Based on $JK$ photometry of 
some 3,300 JAGB stars in the bar of the Large Magellanic Cloud (LMC)
we find that these very red AGB stars have a constant absolute magnitude of $<M_J> 
= -6.22$~mag, 
adopting the Detached Eclipsing Binary (DEB) distance to the LMC of 18.477 $\pm$ 0.004 
(stat) $\pm$ 0.026 (sys). 
Undertaking a second, independent calibration in the SMC, which also has a DEB geometric
distance, 
we find $<M_J> = -6.18 \pm $ 0.01 (stat) $\pm$ 0.05~(sys)~mag. The scatter is 
$\pm$0.27~mag for single-epoch observations, 
(falling to $\pm$0.15~mag for multiple observations averaged over a window of more than 
one year). 
We provisionally adopt $<M_J> = -6.20$~mag $\pm$ 0.01~(stat) $\pm$ 0.04~(sys)~mag for the 
mean absolute magnitude
of JAGB stars. Applying this calibration to JAGB stars recently observed in the 
galaxy NGC~253, 
we determine a distance modulus of 27.66 $\pm$~0.01(stat) $\pm$~0.04~mag (syst), 
corresponding 
to a distance of 3.40 $\pm$~0.06~Mpc~(stat). This is in excellent agreement with the 
averaged TRGB 
distance modulus of 27.68 $\pm$~0.05~mag, assuming $M_I =$ -4.05~mag for the TRGB zero point.

\end{abstract}

.

\medskip
\vfill\eject

\section{Introduction}
$J$-Type Asymptotic Giant Branch (JAGB) stars were first isolated as a
distinct class of objects by Weinberg \& Nikolaev (2001).
Figure 1 reproduces their $K$ vs $(J-K)$ color-magnitude diagram defining eleven distinct 
sequences of stars, as seen in the 2MASS data for the Large Magellanic
Cloud. JAGB stars are high-luminosity asymptotic giant branch (AGB)
stars, whose colors are very red  and whose intrinsic
luminosities are  well defined, and restricted to a narrow
color range. JAGB stars almost certainly contain a subset of the
spectroscopically classified ``carbon-rich" population of
intermediate-age, intermediate-mass stars in galaxies. 
And they also overlap with a subset of Mira variables in specific, 
and with Long-Period Variables (LPV) more generally, with those 
stars being sub-classified as such by their variability, amplitudes, 
colors  and periods, etc. Within the context of this work on the 
extragalactic distance scale and within the following series of papers,
JAGB stars are being defined and identified in terms of purely
single-epoch photometric criteria. Their overlap with 
carbon stars and LPVs will play an interesting, but background role in 
understanding the JAGB stars' evolutionary status, and in optimizing
their practical use as extragalactic distance indicators; but, we
emphasize from the outset that {\it we do not rely either 
on spectral data nor on time-domain information to make efficient use 
of JAGB stars as high-precision and accurate extragalactic distance
indicators.}

In a companion paper (Freedman \& Madore 2020) we discuss the historical
context and empirical evidence for JAGB stars being high-precision distance
indicators. Based on previously published near-infrared data, we extend
our application of this method to a sample of 14 galaxies, both within
the Local Group and out to a distance of 4~Mpc. With this sample
we can assess the accuracy and current precision of the JAGB method by
making a comparison with published distances that are based on the 
tip of the red giant branch (TRGB) method.

In this contribution we use three galaxies to build, calibrate 
and then test this new parallel path in the establishing the 
extragalactic distance scale. The LMC and SMC are used to establish 
the zero-point of the JAGB method, and the galaxy NGC~253 is used 
as an extreme-case system, having been observed at the current distance limit, 
for ground-based telescopes, of about 4~Mpc. Published, near-infrared
color-magnitude diagrams (CMDs) show that JAGB stars have been found 
in abundance and easily measured in all three galaxies.

We conclude that the JAGB method can take its place shoulder-to-shoulder with the Leavitt Law 
for Population~I Cepheids and the TRGB method for Population~II red giant branch stars. Taken 
as a whole, we refer to this tripartite ensemble of stellar distance indicators as forming the 
basis of the {\it Astrophysical Distance Scale.}

\section{The LMC}
Near-infrared $JHK$ data used here for the LMC calibration of the JAGB method 
were taken from Macri et al. (2015). These observations are to be preferred over 
previously-available $2MASS$ data (as used by Weinberg \& Nikolaev 2001) for a number 
of reasons: the Macri photometry is of higher precision and of better spatial resolution 
(alleviating crowding), and most importantly these data are temporally averaged. 
The Macri data, which are all on the 2MASS photometric system, were collected over 13 months, 
resulting in an average of 16 epochs being 
obtained for each star in the survey. The cadence and window over which the observations were made 
were chosen so as to be able to identify known Cepheids and to discover new long-period variables 
in order to determine their periods, amplitudes phases and mean magnitudes in the near infrared. 
The Cepheids are treated in Macri et al., while the Miras are discussed in Yuan et al (2017).

\begin{figure*} 
\centering 
\includegraphics[width=20.0cm, angle=-0]
{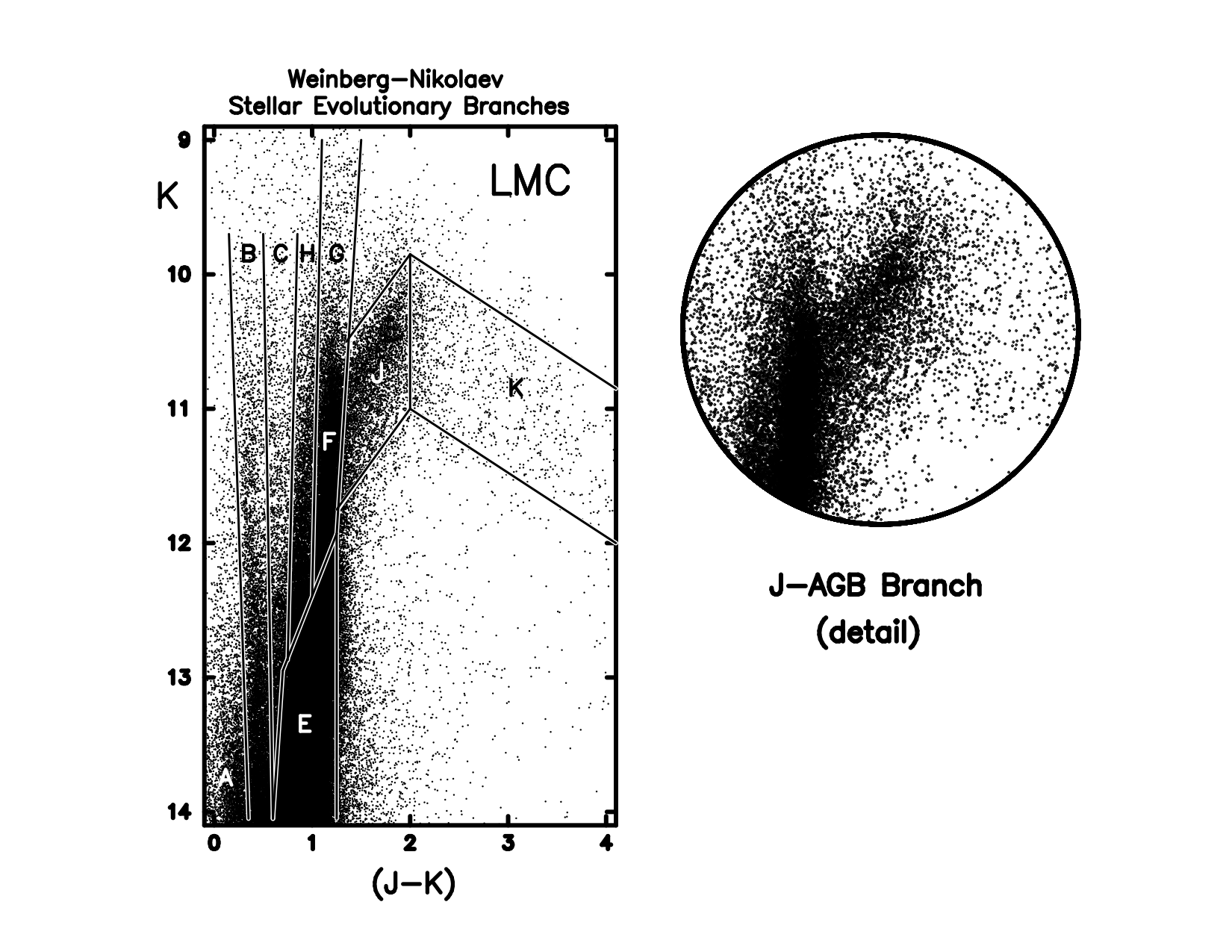} \caption{\small  A near-infrared CMD for 
stars observed by Macri et al. (2015) in the bar of the Large Magellanic Cloud. 
We draw attention to the narrowly defined feature in the center of the plot, 
labeled as $Region ~J$, and highlight this region in the expanded plot to the right. 
This is the defining plot for our JAGB stars. Extreme carbon stars (having developed 
strong winds and probably producing dust in them) sparsely occupy $Region ~K$, 
redward of the JAGB population. Interested readers are referred to the originating 
paper for details on the other branches and populations delimited in this figure,
noting that all other populations are blueward of the JAGB stars studied here.
} 
\label{fig:LMC}
\end{figure*}

It is also important to note that the Macri sample is  confined to 
the bar of the LMC. 
This region is co-located with the Detached Eclipsing Binary ($DEB$) stars, 
used for a geometric determination of the distance to the LMC, were observed by
Pietrzynski et al. (2019). Having the $DEB$s and JAGB stars located
co-spatially minimizes concern that the two populations might be at
different mean distances due to inclination effects and/or the generic
back-to-front geometry of the LMC (e.g., Welch et al. 1987, Weinberg 
\& Nikolaev, 2011, Scowcroft et al. 2011). For the LMC bar and for our JAGB stars 
we adopt a  true distance modulus of $\mu_o = $ 18.477~mag, carrying 
a systematic error of $\pm$0.026~mag, as attributed to the DEB population by Pietrzynski et al.

\begin{figure*} \centering \includegraphics[width=20.0cm, angle=0]
{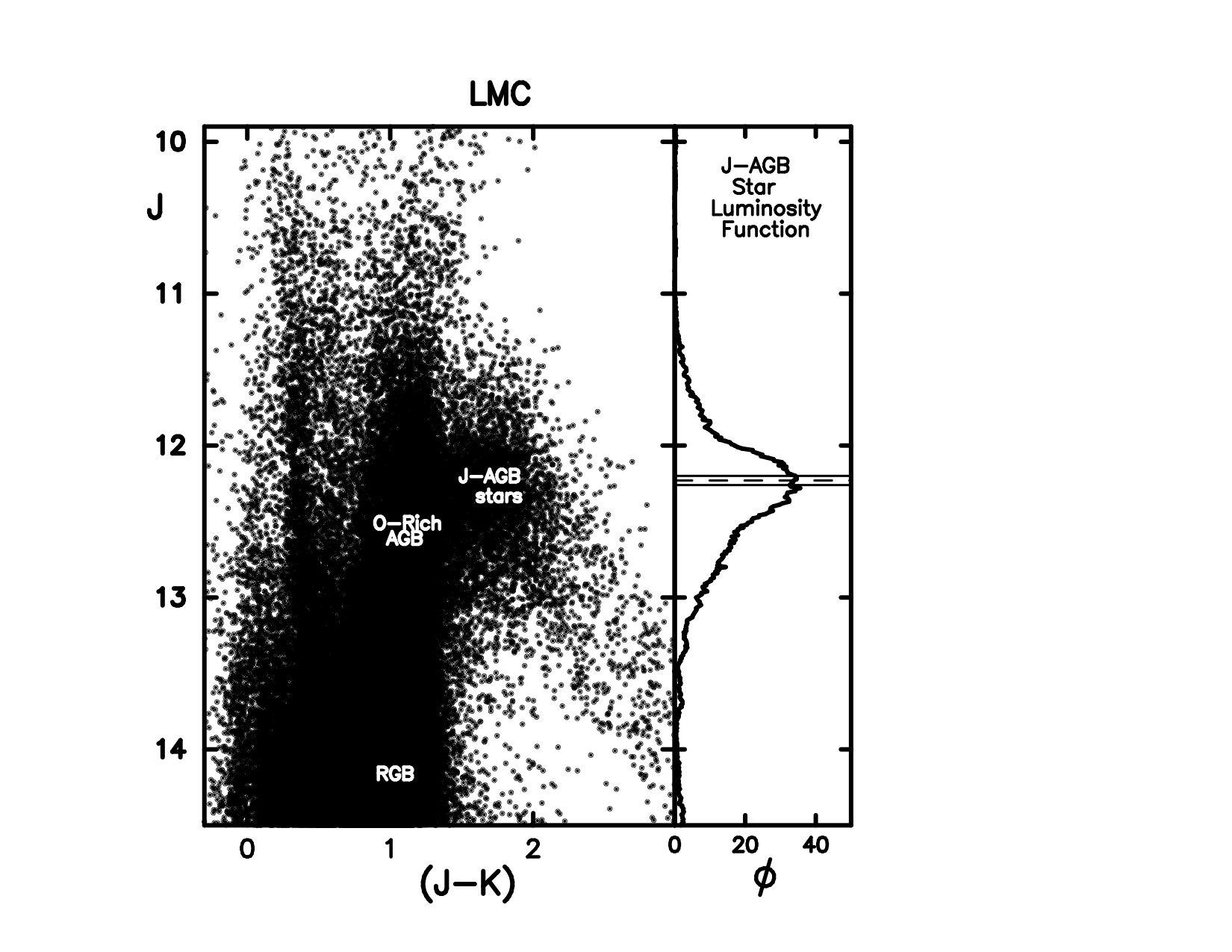} \caption{\small Left Panel: The $J$ vs $(J-K)$ CMD of the LMC 
based on data from Macri et al. (2015). The AGB population, as marked, is seen to be 
constant in luminosity as a function of color within the color range $1.30 \le (J-K) 
\le 2.00$~mag. In the right-hand panel the smoothed luminosity function of stars 
within that same color range is shown.
} 
\label{fig:macri}
\end{figure*}

Figure \ref{fig:macri} shows the $J-(J-K)$ time-averaged, near-infrared CMD for the 3.5 million  
LMC stars published 
in Macri et al. (2015). The dominant population of red stars, rising vertically 
from $J \sim$ 14.5~mag at around $(J-K) =$ 1.0~mag consists of the Population~II
red giant branch (RGB), continuing up to its ``tip" at $J \sim$ 13.5~mag (Hoyt et al. 2018).
This is followed (almost without interruption in this high-density plot) by the 
``oxygen-rich" AGB plume (marked midway at $J \sim$ 12.5~mag). This feature continues up another 
several magnitudes to $J \sim$ 10.5~mag in this diagram. It is important to note that both of these 
red populations, and almost all of the other dominant stellar populations in the near-infrared 
CMD are confined to colors that are {\it bluer} than $(J-K) \sim$ 1.3~mag.  
Redward of this limit there is only one noteworthy population, that being the object 
of this study, {\it i.e.,} the population of JAGB stars.

Operationally, we simply define JAGB stars to have near-infrared colors between 1.30 $ < 
(J-K) <  $ 2.00~mag. The red limit eliminates the possibility of low level contamination from 
extreme carbon stars that fall in magnitude and continue out to $(J-K) >$ 4.5~mag, 
constituting what Weinberg \& Nikolaev (2001) have independently designated as 
constituting a separate population, the ``$K$ Branch" (see Figure \ref{fig:LMC}).

\section{Absolute Calibration}

As we proceed in the calibration of the absolute magnitude of the JAGB method we begin by 
reviewing the conclusions reached earlier by Weinberg \& Nikolaev (2001). They found a tight 
correlation between the $(J-K)$ color and the $K-$band luminosity of the ``$J$-branch" stars 
as delimited and separated out in Figure \ref{fig:LMC}. A luminosity-color fit to their 2MASS data in 
the $J$ Branch gave $ K = D_o - 0.99 (J-K) $, where $D_o$ is the $K$-band zero point of 
this particular ``color-magnitude" relation. What Weinberg \& Nikolaev did not comment on 
was that, by (not unreasonably) adopting a unit slope for the color term, their relation
becomes 
$$ K~ ~= D_o ~-~ (J-K) $$ 
which, upon transposing the color term, becomes
$$J = K + (J-K)~ =  D_o$$
or, more to the point 
$$J =  constant $$ Simply stated, in the near-infrared $J$-band, JAGB stars have a constant 
mean magnitude, independent of color. 

Weinberg \& Nikolaev (2000) measured the scatter in their single-epoch data to be on 
the order of $\sigma_{JAGB} =  \pm$0.33~mag. However, in the presence of known variability 
in the JAGB star population, the question naturally arises as to what the intrinsic 
(i.e., the time-averaged) scatter is for these stars. If we assume that the 
average amplitude of these stars is 0.7~mag (which corresponds to an equivalent sigma 
of $\pm$0.22~mag) then subtracting this in quadrature from the observed (single-epoch) 
scatter $\pm$0.33~mag  gives an {\it upper limit} of $\sigma_o = \pm $0.25~mag for 
the intrinsic scatter of the JAGB method. The observed, time-averaged
scatter seen in the JAGB LMC-bar sample (Macri et al.) is found to be
$\pm$0.15~mag. Given the possibility of additional, currently
unaccounted for, sources of scatter (residual depth effects, crowding 
and photometric errors, etc.) we take this, already small, value for 
the intrinsic dispersion to still be an upper limit on the method.\footnote{
Indeed, some Mira are known to have amplitudes up to 10 mag
(peak-to-peak) in the blue.  However, most of what is (residually)
seen in the NIR is the sinusoidal variation of the radius, giving
$K$-band amplitudes of less than 0.5 mag according to Weinberg \& Nikolaev. Similarly,
Smith et al. (2002) cite average $J$-band amplitudes of 0.5~mag. On the
other hand Huang et al. (2018) find $H$-band amplitudes of LMC Miras
typically fall in the range of 0.4 to 1.3 mag, with an average
amplitude of about 0.8~mag. Here we have adopted a provisional value of
0.7 mag for the average $J$-band amplitude).
Weinberg \& Nikolaev (2000) give a detailed break down of the various
additional terms contributing to their single-phase observed scatter of
$\pm$0.33~mag. Those factors include back-to-front depth, photometric
errors, differential extinction, and intrinsic scatter in their $J$
Branch mean stellar magnitudes, and in their own words, they conclude that
``This is direct evidence that our color-selected sources are standard
candles at least as good as $\sigma \sim $ 0.2~mag. In fact, they are even
better ..."} .

\subsection{The Large Magellanic Cloud (LMC)}
By adopting a true distance modulus to the LMC bar of $\mu_o = $~18.477~mag 
and an apparent mean magnitude of $J = $12.31~$\pm$~0.01~mag for the 3,341 
JAGB stars studied here (also in the bar), 
we find that the mean absolute $J$-band zero point of the JAGB method 
(corrected for $A_J = $~0.053~mag of foreground extinction, NED) is
$M_{J_o}(LMC) = $ -6.22~mag $\pm$ 0.01 ~(stat) $\pm$ 0.03 ~(syst).

\vfill\eject

\subsection{The Small Magellanic Cloud (SMC)}
The SMC may not seem to be an ideal object for testing the 
JAGB method, given that it is well known to be highly disturbed and tidally 
extended on the sky (e.g., Scowcroft et al 2017, and references therein), presumably 
because of interactions with both the LMC and the Milky Way.
But, one can select a sub-sample of JAGB stars found in a one-degree-radius
circular region centered on the main body of the SMC, which may arguably be less dispersed
along that particular line of sight, when compared to the tidal tails, far from the main 
body (see Scowcroft et al.). The $JHK$ photometry for that subset of 116,298 SMC  stars 
was extracted from the 2MASS on-line version of that survey, available through {\it IRSA.}

\begin{figure}[htb!] \includegraphics[width=20.5cm, angle=-0]
{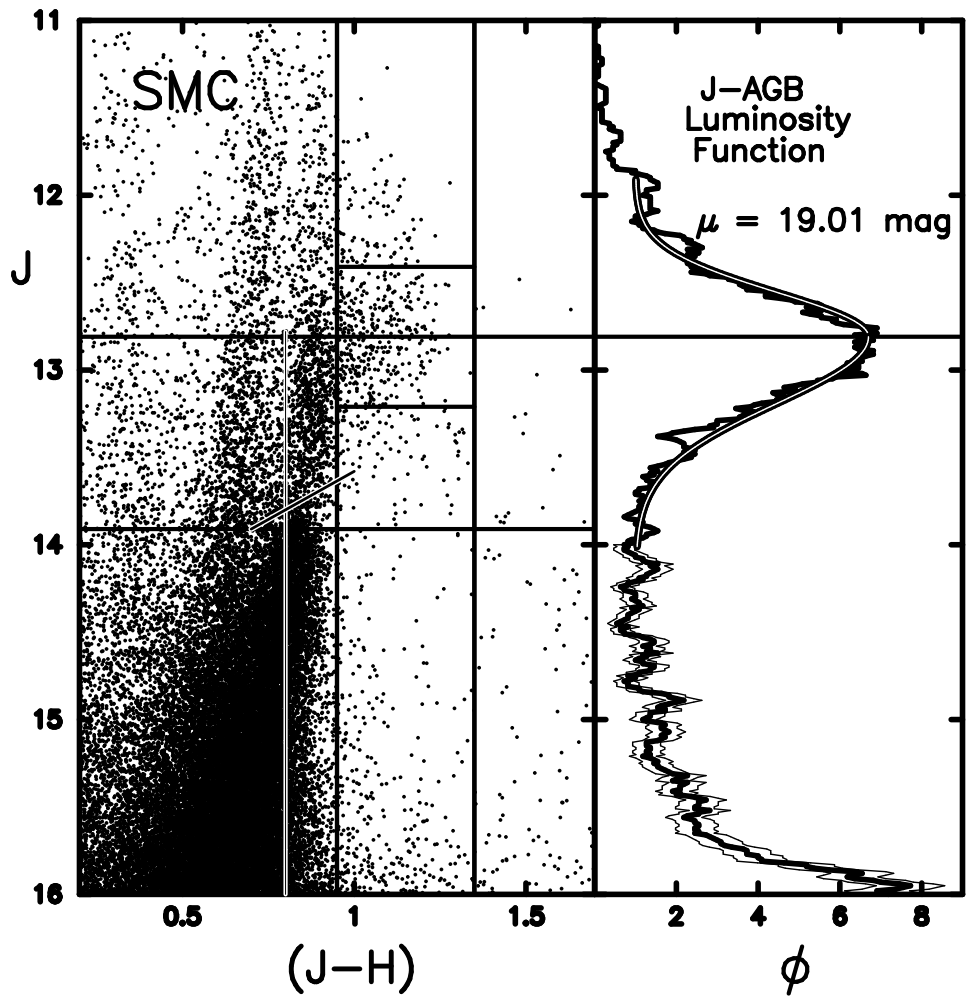} \caption{\small  CMD and Luminosity Function for 3,405 luminous red (JAGB) 
stars in the Small Magellanic Cloud. See Figure \ref{fig:macri} and text for details. } 
\label{fig:SMC}
\end{figure}

The CMD and luminosity function for the JAGB stars in the SMC are each shown in Figure \ref{fig:SMC}. 
3,405 JAGB stars contributed to the color-selected luminosity function, yielding a mean apparent magnitude of $J = 12.81\pm$0.01~mag. The DEB distance modulus to the SMC is $\mu_o = $~ 18.965~mag,
as published by Graczyk et al. (2014). Those DEB stars are centered on this same region of the SMC, making it not unreasonable to assume that their mean distance is the same as the average distance of the selected JAGB stars. Applying a foreground Galactic extinction correction of $A_J = $ 0.026~mag to the apparent magnitude of the JAGB stars, and correcting for distance yields $M_{J_o}(SMC) $=  -6.18~mag $\pm$ 0.01 ~(stat) $\pm$ 0.05 ~(syst).

\section{A Provisionally Adopted JAGB Zero Point}

Taking the two geometrically-calibrated zero points, based on DEB distances to the LMC and SMC we arrive at our first estimate of the $J$-band zero point of the JAGB distance scale:
$M_{Jo} = $-6.20~mag $\pm$ 0.01~(stat) $\pm$ 0.04~(syst). 

\medskip

\par
For single-epoch observations, the scatter in the JAGB method is found by Weinberg \& Nikolaev (2001) to 
be $\pm$0.27~mag. For time-averaged observations, the scatter reduces to $\pm$0.15~mag (Macri et al. 2015). 
In conservatively adopting the larger dispersion, a sample of 200 JAGB stars can deliver a distance 
that is statistically good to a precision of 1\%. To put that sample size in perspective, even in the 
restricted bar region of the LMC (itself a relatively low-mass, 
low-luminosity galaxy) there are over 3,000 of these stars cataloged. 

All galaxies with intermediate-age populations are now candidates for 
having high-precision distances determined by the JAGB method. 
In what follows, we make the first application of this method to the galaxy 
NGC~253 at 3.5~Mpc, some 70 times more distant than the LMC. 

\begin{figure}[htb!] 
\includegraphics[width=16.5cm, angle=0]
{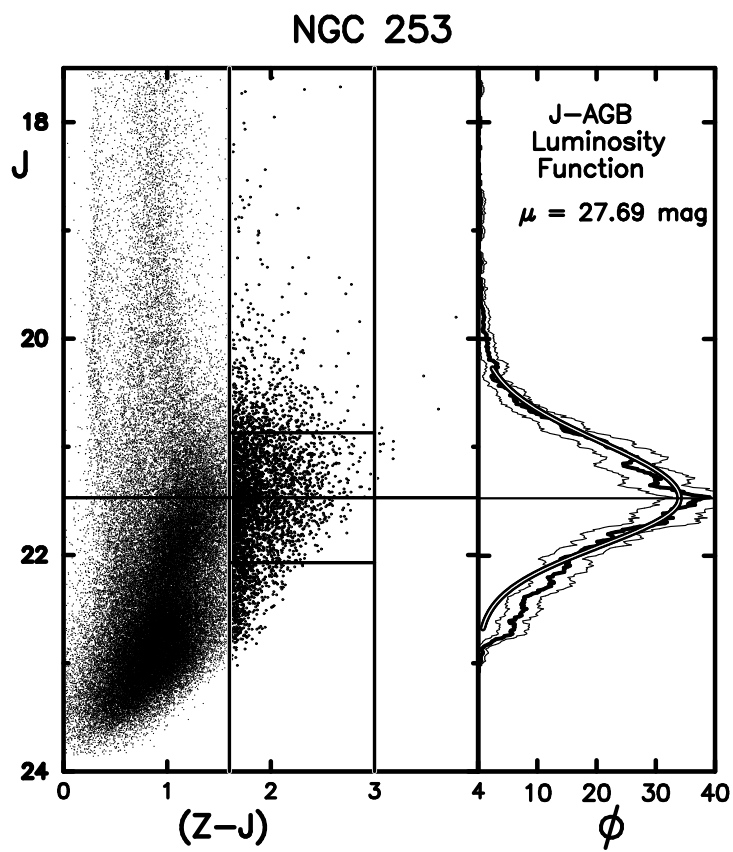} \caption{\small CMD (left) and Luminosity Function (right) for luminous red stars in 
NGC~253.} 
\label{fig:N253}
\end{figure}

\section{NGC 253: A First Test of the JAGB Method Outside of the Local Group}
For NGC 253 there are four modern HST detections of TRGB stars outside of the 
main disk of this nearly edge-on galaxy. Each determination has been standardized
to an absolute $I$-band magnitude for tip of $M_I = $ -4.05~mag and uniformly
corrected for the same foreground extinction of $A_I = $ 0.136~mag (NED).
The homogenized distance moduli are 27.53~mag (Mouchine et al. 2005), 
27.68~mag (Dalcanton et al. 2009), 27.67~mag (Rayburn-Smith et al. 2011) and
27.82~mag (Jacobs et al. 2009). The averaged TRGB distance modulus is found 
to be $<\mu_o (TRGB)> = $~27.68~mag $\pm$ 0.06 (scatter on the mean).

Ground-based $Z$ and $J$ band data, used for the JAGB detection, were published 
by Greggio et al. (2014) based on observations made as part of the $VISTA$ science 
verification process. The CMD is shown in Figure \ref{fig:N253}.
3,169 JAGB stars in NGC~253 give an apparent distance modulus
of $\mu_J (JAGB)$ = 27.67 $\pm$ 0.006~mag. Correcting for a foreground
extinction of $A_J =$ 0.013~mag (NED), the true JAGB 
distance modulus to NGC~253 then becomes $\mu_o (JAGB) = $ 27.66~$\pm$ 0.01 mag (stat).

The agreement of these two determinations, made at the current limit of ground-based observing, 
is very encouraging, and foreshadows the confirming results of an inter-comparison of a 
significantly larger sample of JAGB and TRGB distances to nearby galaxies, given in 
Freedman \& Madore (2020). But what can be said here is that the low-metallicity 
calibration made using the LMC [Z $=$ 8.50] amd SMC [Z $=$ 7.98] appears to apply equally 
well to the high-metallicity galaxy NGC~253 [Z $=$ 8.99, (Zaritsky et al. 2000, which is 
almost identical to the metallicity of M~31, at Z $=$ 8.98] whose JAGB distance 
is in exact agreement with its independently-determined TRGB distance.

Before concluding this paper, we next give a short
description of the physics behind the JAGB method, 
showing that current models can justifiably claim 
to describe most of the salient aspects of this promising,
new distance indicator. 

\vfill\eject
\section{A Very Brief Introduction to Stellar Evolution Models}
\begin{figure}[htb!] 
\includegraphics[width=14.5cm, angle=0]
{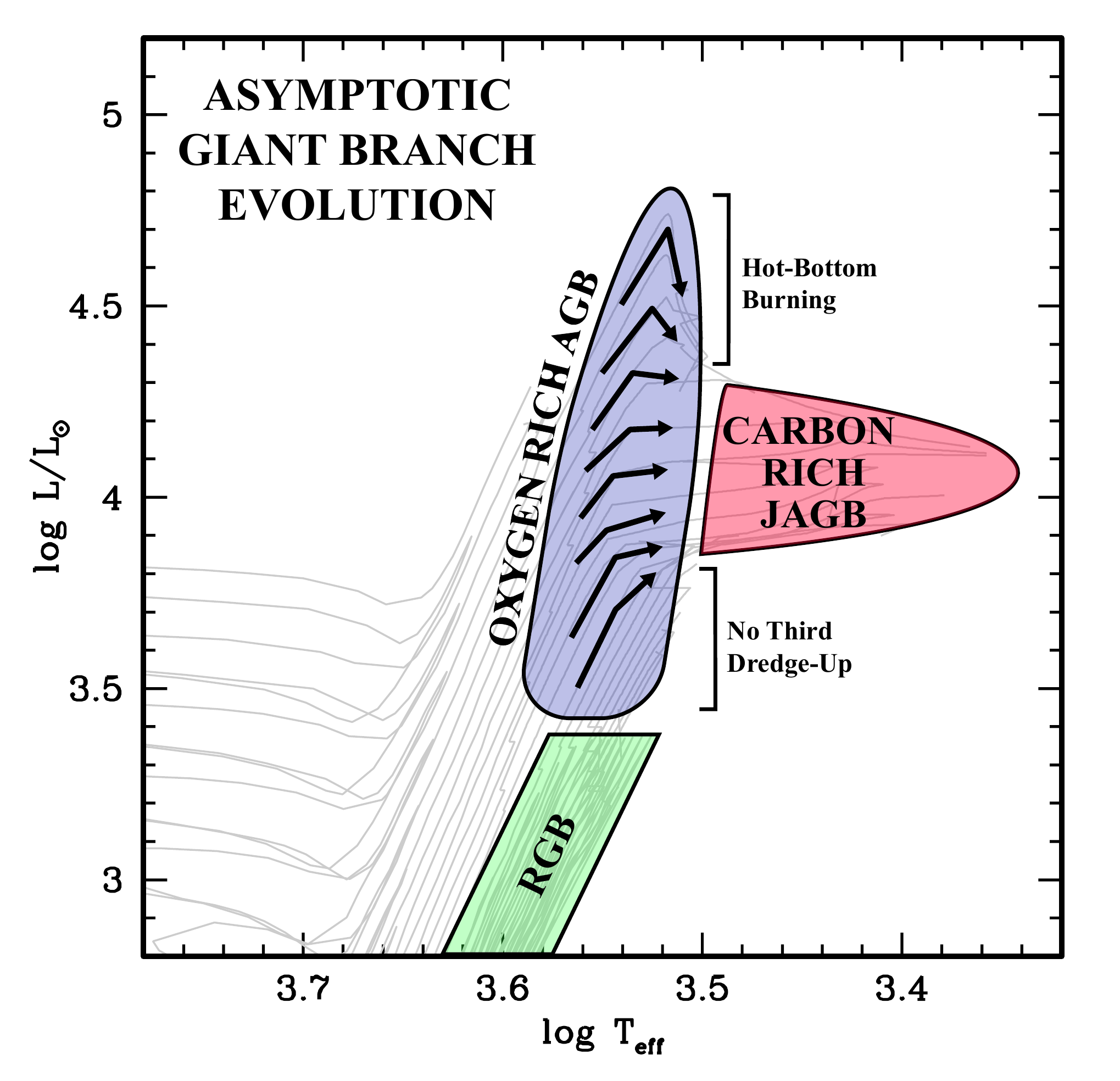} \caption{\small Theoretical models (in light blue)  from Marigo et al. 
(2008) are plotted in the $log T_{eff}$ versus $log L$ plane, showing the final phases 
of asymptotic giant branch evolution leading to the carbon-rich JAGB stars. The first-ascent 
RGB stars are highlighted in green, and the oxygen-rich AGB stars are shown in blue. The 
region containing the carbon-rich JAG stars is highlighted in red. The funneling effect is
the result of (1) ``Hot-Bottom Burning", at high luminosities, destroying the carbon before it 
can be brought to the surface, and then (2) at lower luminosities and lower masses, 
``the third dredge-up phenomenon" no longer occurs, and the earlier two dredging events do not 
reach deep enough to touch the shell-burning carbon-rich region. These two cut-off regions are 
marked and named above and below the ``Carbon Rich" JAGB region.}

\label{fig:cartoon}
\end{figure}

Stellar evolution models have become sufficiently sophisticated of late that they
can accommodate many of the details leading up to and including the transition
producing  carbon-rich AGB stars from their thermally-pulsing (TP) oxygen-rich
precursors (see Habing \& Olofsson 2004 for extensive reviews, and especially 
Marigo et al. 2008, 2017 for more recent updates.) The intermediate-age,
intermediate-metallicity TP-AGB stars, that are of interest to 
us here, as a whole have a narrowly-defined range of masses, going from 2 to 5~M$_\sun$. 
As Marigo et al. (2008) show (and as can be seen in Figure 5), these same stars (that have ages 
between 300 Myr and 1 Gyr) are also tightly constrained in luminosity at their latest stages,
being funneled into a carbon-rich phase whose {\it total range} is only $\Delta log(L/L_{\sun})$ =  
0.5 (i.e., $\sigma = \pm$0.31~mag)  at Z = 0.008. Those evolutionary tracks are 
shown in Figure \ref{fig:cartoon} as a muted backdrop to a schematic of the 
RGB/AGB populations highlighted in color, and emphasizing a variety of interesting 
features: (1) the RGB population (in green) terminates at its so-called
`tip' (horizontally defined at $log(L/L_\sun)$ =  3.4) at a luminosity that is 
at least 3 magnitudes fainter than the upper limit of oxygen-rich AGB population,
rising above the TRGB, at approximately the same color and effective temperature. 
Bold black arrows trace the evolutionary tracks within the oxygen-rich AGB
phase, showing the focusing of those tracks on the carbon-rich phase at significantly 
lower effective temperatures, and within the aforementioned small range of luminosities. 

It should be noted that the stellar interior physics, predicting the narrowly
confined luminosities of carbon stars has been part of stellar evolution theory
for nearly half a century. As for the reaction of the atmospheres, that situation
is succinctly summarized by Marigo (2008) wherein ``... the main physical effects
driving the appearance of the red tail of C stars, are the cool effective
temperatures, caused by changes in molecular opacities, as the third dredge-up 
events increase the C/O ratio."

The reasons for the upper and lower cut-offs to the oxygen-rich AGB evolution
towards the red have also been known and understood for decades. The sole 
reason for these advanced-evolutionary-phase stars becoming as red as some of
them do is because of carbon being transferred from the interior out into the
envelope (and then into the atmosphere). The ``dredge-up" episodes (number 3 and 
beyond), that penetrate deep enough into the interior to reach the carbon-rich, 
helium-burning shell, are only effective in bringing carbon to the surface
for masses in excess of 1.2-1.3~$M_{\sun}$ (Groenewegan \& Marigo 2004). Below 
that critical mass AGB stars cannot, and do not, develop carbon-rich 
atmospheres. This leads to a {\it lower bound} on carbon 
star luminosities (Iben \& Renzini 1983). At the other extreme, an {\it upper 
limit} on carbon star masses (and luminosities) comes from the destruction of 
carbon at the base of the convective zone. This occurs when the temperature 
in the carbon-rich interior region becomes so hot that carbon itself burns, 
leaving none of it left to be convected to the surface. This phase transition to
``Hot-Bottom Burning" turns on for AGB stellar masses in excess of 
3.5~$M_{\sun}$ (Iben 1973, Sackmann et al. 1974), and it may be cutting in at
2$M_{\sun}$ for very low metallicity stars (Siess et al. 2002).  Finally, 
Karakas (2014), in her Figure 3, shows that the distribution functions of
the C/O ratio as a function of initial mass, all come to a maximum within the
narrow mass range of 3.0 to 3.6 $M_{\sun}$, for the three metallicities
she considers (those being Z $=$ 0.007, 0.014 \& 0.030). Her models also indicate
that C/O ratios on the surface exceed unity predominantly in the mass
range from 2.0 to 4.5 $M_{\sun}$.

\section{Conclusions}

JAGB stars are easily identified in the near infrared by their extremely red colors.
They are also very luminous and have a well-defined mean magnitude in 
the $J$ band that is constant with color. Based on a sample of 3,300 JAGB
stars in the bar of the LMC, and a similar number of JAGB stars in the main body 
of the SMC, we have calibrated the mean absolute magnitude of JAGB stars 
in the near-infrared $J$ band. Using the detached eclipsing binary (DEB) distance 
modulus of 18.477~mag to the LMC and an SMC DEB distance modulus of 18.965~mag, 
we provisionally find the mean $J$-band absolute magnitude of the JAGB stars to be  
$M_{J_o} = $ -6.20 $\pm$~0.01~(stat) $\pm$~0.04~(sys)~mag. Applying this 
calibration to NGC~253, we derive a JAGB distance modulus of 27.68 $\pm$~0.01~(stat) 
$\pm$~0.04~(sys)~mag, which corresponds to a metric distance of 3.44 $\pm$~0.06~Mpc.

A more extensive application of the JAGB method to galaxies within and beyond the Local 
Group, extending out to 4~Mpc is presented in a companion paper (Freedman \& Madore 2020).


\section{Acknowledgements}  We thank the {it University of Chicago} 
and {\it Observatories of the Carnegie Institution for Science} 
and the {\it University of Chicago} for their support of 
our long-term research into the calibration and determination 
of the expansion rate of the Universe. Support for this work was 
also provided in part by NASA through grant number HST-GO-13691.003-A 
from the Space Telescope Science Institute, which is operated by AURA,
Inc., under NASA contract NAS~5-26555. This research would 
simply not have been possible without the NASA/IPAC Extragalactic 
Database (NED), which is operated by the Jet Propulsion Laboratory,
California Institute of Technology, under contract with the 
National Aeronautics and Space Administration.

\vfill\eject
\section{References}

\medskip

\noindent
Dalcanton, J.J., Williams, B.F., Seth, A.C., et al. 2009, ApJS, 183, 67

\noindent
Freedman, W.L. \& Madore, B.F. 2020, ApJ, submitted

\noindent
Graczyk, D., Pietrzynski, G., Thompson. I., et al. 2014, ApJ, 780, 59

\noindent
Greggio, L., Rejkuba, M., Gonzales, O.A., et al.  2014, A\&Ap, 562, 73

\noindent
Groenewegen, M.~A.~T. \& Marigo, P. 2004 ``Asymptotic Giant Branch Stars'',  eds. H.~J. 
Habing \& Olofsson, H. Springer-Verlag, New York, pg 105

\noindent
Habing, H.~J. \& Olofsson, H. 2004, ``Asymptotic Giant Branch Stars'',  Springer-Verlag, New York

\noindent
Hoyt, T., Freedman, W.L., Madore , B.F., et al. 2018, ApJ, 858, 12

\noindent
Huang, C.D., Riess, A.G., Hoffmann, S.L. 2018, ApJ, 857, 67

\noindent
Iben, I. 1973, ApJ, 185, 209

\noindent
Iben, I. \& Renzini, A. 1983, ARAA, 21, 271

\noindent
Jacobs, B.A., Rizzi, L., Tully, R.B., et al. 2009, AJ, 138, 332

\noindent
Karakas, A.I. 2014, MNRAS, 445, 347

\noindent
Macri, L., Ngeow, C.-C., Kanbur, S., Mahzooni, S., \& Smitka, M. T. 2015, AJ, 149, 117

\noindent
Marigo, P., Girardi, L., Bressan, A., et al. 2008, A\&A, 482, 883

\noindent
Marigo, P., Girardi, L., Bressan, A., et al. 2017, ApJ, 835, 77

\noindent
Mouchine, M., Ferguson, H.C., Rich, R.M., et al.  et al. 2005, ApJ, 633, 810

\noindent
Pietrzynski, G., Graczyk, D., Gallenne, A., et al., 2019, Nature, 567, 200
 
\noindent
Radburn-Smirh, D.~J., de Jong, R.~S., Seth, A.~C., et al. 2011, AJ, 195, 18

\noindent
Sackmann, I.J., Smith, R.L. \& Despain, K.H. 1974, ApJ, 187, 555

\noindent
Scowcroft, V., Freedman, W.L., Madore, B.F., et al. 2011, ApJ, 743, 76

\noindent
Seiss, L., Livio, M. \& Lattanzio, J. 2002, ApJ, 570, 329

\noindent
Smith, B.J., Leisawitz, D., Castelaz, M.W  \& Luttermoser, D. 2002, AJ, 123, 948 

\noindent
Weinberg, M. D., \& Nikolaev, S. 2001, ApJ, 548, 712

\noindent
Welch, D.L., McLaren, R.A., Madore, B.F., \& McAlary, C.W. 1987, ApJ, 321, 162

\noindent
Yuan, W., Macri, L., He, S., et al. 2017, ApJ, 154, 149

\noindent
Zaritsky, D., Kennicutt, R.C., \& Huchra, J.p. 2000, ApJ, 420, 87 

\vfill\eject
\end{document}